\documentclass[sigconf]{acmart}

\usepackage{packages}

\setcopyright{acmlicensed}
\copyrightyear{2026}
\acmYear{2026}
\acmDOI{XXXXXXX.XXXXXXX}
\acmConference[CHASE 2026 DECS]{Doctoral and Early Career Symposium of the 19th International Conference on Cooperative and Human Aspects of Software Engineering}{April 13--14,
  2026}{Rio de Janeiro, Brazil}
\acmISBN{978-1-4503-XXXX-X/2026/XX}

\begin{document}

\title{Personalizing LLM-Based Conversational Programming Assistants}

\author{Jonan Richards}
\email{jonan.richards@ru.nl}
\orcid{0009-0007-1218-8599}
\affiliation{%
  \institution{Radboud University}
  \city{Nijmegen}
  \country{The Netherlands}
}

\begin{abstract}
Large Language Models (LLMs) have shown much promise in powering a variety of software engineering (SE) tools. Offering natural language as an intuitive interaction mechanism, LLMs have recently been employed as conversational ``programming assistants'' capable of supporting several SE activities simultaneously. As with any SE tool, it is crucial that these assistants effectively meet developers' needs. Recent studies have shown addressing this challenge is complicated by the variety in developers' needs, and the ambiguous and unbounded nature of conversational interaction. This paper discusses our current and future work towards characterizing how diversity in cognition and organizational context impacts developers' needs, and exploring personalization as a means of improving the inclusivity of LLM-based conversational programming assistants.
\end{abstract}



\maketitle

\section{Introduction}
Large Language Models (LLMs) have shown much promise in performing a variety of code-centric software engineering (SE) tasks such as code generation, bug detection, and in supporting developers' code understanding~\cite{hou2024LargeLanguageModels}. LLM-based tools that are implemented as conversational assistants, hereafter referred to as ``programming assistants'', hold potential for SE by not only providing technical utility but also supporting human aspects of software development. Programming assistants can support several SE activities at once~\cite{ross2023ProgrammersAssistantConversational}, can increase productivity gains by allowing iterative refinement and progress towards the developer's goal~\cite{austin2021ProgramSynthesisLarge}, and is missed by developers when not present~\cite{liang2024LargeScaleSurveyUsability}. Examples of programming assistants include in-IDE chat functionalities of GitHub Copilot\footnote{\url{https://github.com/features/copilot}} and Tabnine\footnote{\url{https://www.tabnine.com/ai-chat/}}, but also encompass more general assistants such as ChatGPT\footnote{\url{https://openai.com/index/chatgpt/}} which are often used by developers in coding contexts~\cite{xiao2024DevGPTStudyingDeveloperChatGPT}.

Despite their apparent potential, several links have been found between challenges in using programming assistants and developers' experience, learning style, and gender~\cite{nam2024,choudhuri2024HowFarAre,nguyen2024,russo2024NavigatingComplexityGenerative}, which may partially explain the gender and experience gap in the adoption of LLM-based tools~\cite{draxler2023GenderAgeTechnology}. While it is well-known that cognitive diversity, which has been linked to gender, affects people's software-based problem-solving styles~\cite{burnett2016GenderMagMethodEvaluating}, this evidence suggests that it also affect developers' needs and behavior when interacting with programming assistants. Besides cognition, organizational factors such as policies, team dynamics, and the position of developers along the software development lifecycle also impact their needs and challenges they run into~\cite{khojah2024}.

LLMs can inherently tailor outputs to personal preferences by allowing for natural language interaction, which developers greatly appreciate~\cite{russo2024NavigatingComplexityGenerative,liang2024LargeScaleSurveyUsability}. However, effectively prompting programming assistants requires substantial effort and experience~\cite{nam2024,nguyen2024,liang2024LargeScaleSurveyUsability}. Clearly, this inherent personalized support has a high barrier to entry, and the challenges faced by only specific groups of developers still highlight a lack of support for cognitive diversity, including experience and problem-solving styles. This line of research is aimed at characterizing the impact of cognitive diversity and organizational factors on individual developers' needs and interaction style regarding programming assistants, and designing new personalization approaches to increase inclusivity of these tools.

\vspace{2pt}
\boxRQ{}{How can an LLM-based programming assistant provide effective personalized support to increase inclusivity to cognitive diversity and organizational context?}

The main contributions of this research will be \begin{enumerate*}[label=\arabic*)]
    \item an explanatory framework of how cognitive and organizational factors affect developers' needs and interaction styles,
    \item the design of personalization strategies to improve inclusivity towards these factors, and
    \item a personalized programming assistant prototype
\end{enumerate*}. Together, these contributions will provide researchers with insights on developer behavior, support developers through tailored interactions, inform programming assistant providers in designing more inclusive support, and help organizations deploy tools that better accommodate diverse developers.

\section{Background}
This research builds on the existing body of human-computer interaction (HCI) research. User experience (UX) is often decomposed into pragmatic UX (e.g. usefulness and usability) and hedonic UX (e.g. enjoyment, autonomy)~\cite{perrig2024MeasurementPracticesUser}. According to Hassenzahl's model of UX~\cite{hassenzahl2003ThingUnderstandingRelationship}, hedonic qualities arise from how well a product enables the user's basic psychological needs and the resulting ``be-goals'', i.e. how the user wants to be or feel (such as in control, knowledgeable, supported). These ``be-goals'' give rise to users' ``do-goals'' (what they want to do using the product), and the extent to which these are enabled by the product results in the pragmatic experience. Norman's Seven Stages of Action model~\cite{norman1986CognitiveEngineering} describes the cyclical process of how these ``do-goals'' are translated into intentions and in turn into actions which are then executed. Next, the interactive system's results are perceived, interpreted and evaluated with respect to the goals and intentions, possibly leading to revised intentions and actions. Subramonyam et al.~\cite{subramonyam2024BridgingGulfEnvisioning} explore challenges users might face in this process when interacting with LLMs, including the setting of goals and intentions so that LLMs can accomplish the task, translating their intentions into optimal actions (prompts), and knowing what to expect from LLMs' output.

It is commonly accepted in HCI research that both personal factors (e.g. cognitive factors, experience with a product, attitude) and contextual factors (e.g. social and organizational context) greatly impact users' needs, behavior, and the challenges they face~\cite{burnett2016GenderMagMethodEvaluating,benyon2019DesigningUserExperience}. Personalization is a commonly used strategy to ensure information systems cater to the wide range of user needs and behavior arising from diversity in personal and contextual factors~\cite{fan2006WhatPersonalizationPerspectives}. However, personalization encompasses a variety of approaches in practice: whereas adaptive systems employ implicit personalization to user behavior and user needs inferred from this behavior, adaptable systems employ explicit personalization by allowing users to configure the system to their own needs~\cite{fan2006WhatPersonalizationPerspectives}.

\section{Research Overview}
\begin{figure}[ht]
  \centering
  \includegraphics[width=\linewidth]{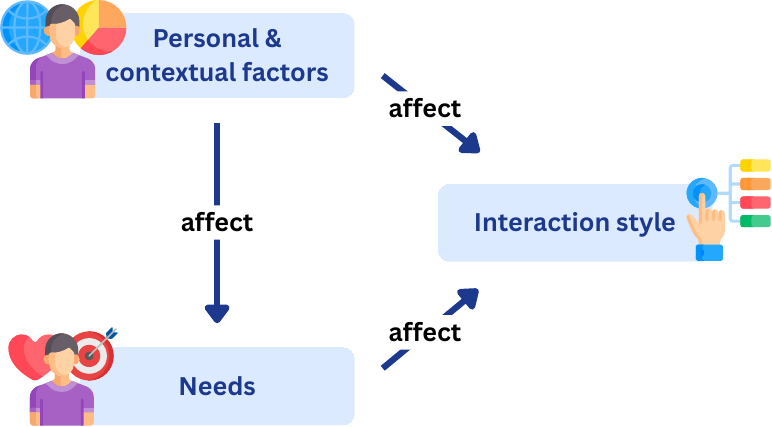}
  \caption{Conceptual model of diversity in interaction used in this research.}
  \label{fig:concept}
\end{figure}

In this research, I refer to developers' psychological needs and their ``be-goals'' as their \textit{needs}, and to the cyclical process of transforming goals into intentions and actions (i.e. behavior) as their \textit{interaction style}. The resulting relationships between these concepts and personal and contextual factors are shown in figure~\ref{fig:concept}. As a hypothetical example, take a junior developer, new to a company, using a programming assistant. The personal and contextual factors include them having relatively little programming experience, and using the assistant in an on-boarding context. These factors could affect their needs, prioritizing their need for becoming familiar the code base over the need for productivity. This need for learning may induce the developer to adopt an interaction style where they ask the assistant to explain the code base and give pointers to a solution to the task at hand, rather than just provide the solution right away. The developer's personal factors may influence their interaction style as well, as their low programming experience could lead them to phrase their questions in a more abstract and high-level way than an experienced developer might.

\begin{figure*}[tb]
  \centering
  \includegraphics[width=\textwidth]{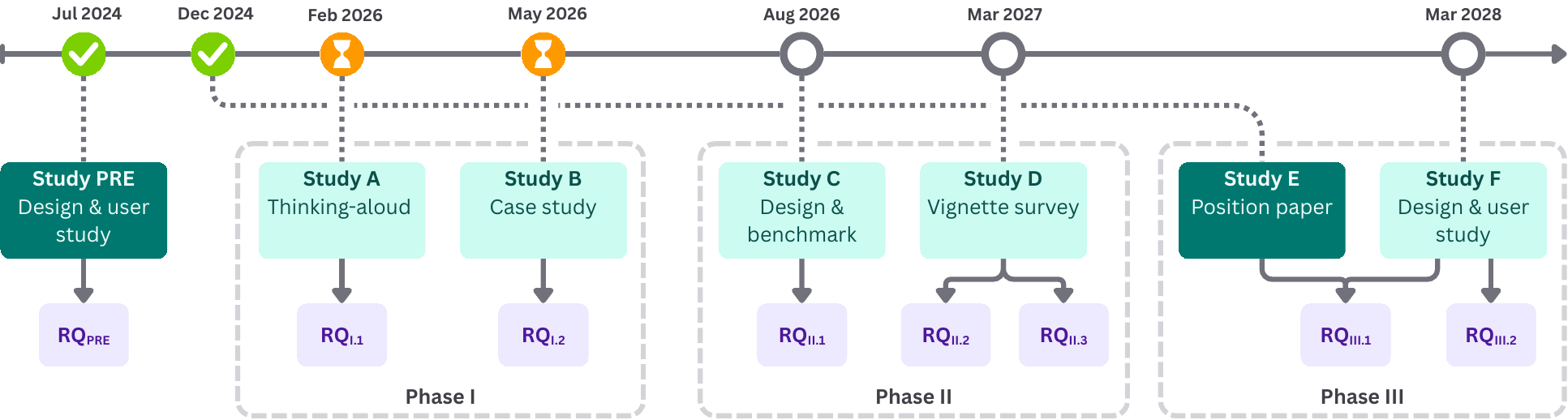}
  \caption{Overview of research phases, including a timeline for completion.}
  \label{fig:overview}
\end{figure*}

To improve both pragmatic and hedonic UX, a personalized programming assistant will need to take into account developers' needs, and accommodate developers' challenges in interacting by considering their interaction style. A developer's interaction style is directly observable to a programming assistant, but their needs will need to be inferred from their interaction style or specified through human-in-the loop methods. We will explore both these implicit and explicit approaches. Figure~\ref{fig:overview} shows an overview of the planned phases for this research, from understanding diversity in interactions (phase I), to designing personalization approaches (phase II) and implementing and evaluating a prototype assistant (phase III).

Since the research involves human subjects, ethical approval will be obtained from the relevant ethics board before each phase.

\subsection{Preliminary Study}

\boxRQ{PRE}{Is automatic personalization of programming assistants feasible?}

To assess the feasibility of fully automatic personalization and answer \refRQ{PRE}, we ran a preliminary study, \boxStudy{PRE}~\cite{richards2024WhatYouNeed}. We decided to initially focus only on personal factors by fixing the context of the experiment to a code understanding setting. In this study, we designed a code understanding assistant prototype that infers developers' needs, leveraging established prompting techniques to improve LLMs' Theory of Mind (the ability to reason about others' mental states). Due to the exploratory nature of the study, no explicit guidance was given to the LLM on the relevant aspects of developers' mental state, or on how to use the inferred needs to adapt the support accordingly.

We evaluated the approach by comparing it to a non-personalized assistant prototype in a within-subject study with fourteen novice developers to capture their perceptions and preferences. Personalization was perceived to improve the assistant's understanding of novices' queries and the clarity of its responses. We also found that participants' interaction styles with the assistants varied significantly, showing that even the constrained participant sample of novices exhibited great diversity. Furthermore, different interaction styles were shown to significantly impact the personalized assistant's effectiveness. Although we demonstrated the feasibility of personalizing a code understanding assistant, it is clear LLMs need explicit guidance on how to personalize. To provide this guidance, we first need a thorough understanding of the interplay between developers' personal and contextual factors, needs, and interaction style.

\subsection{Phase I -- Understanding Diversity in Programming Assistant Interactions}

\boxRQ{I.1}{How does experience and cognitive style affect developers' needs and interaction style in programming assistant interactions?}
\boxRQ{I.2}{How does organizational context affect developers' needs and interaction style?}

In the initial phase of this research, we attempt to understand the effect diversity has on developers' interactions with programming assistants. Although intersectionality likely plays a role, we decided to study the impact of personal and contextual factors separately. This keeps methodological complexity within feasible limits, while allowing us to gain insights into the extent and patterns of diversity in interactions, as well as better ways to support it.

Although there is preliminary evidence of usability issues in programming assistants faced by certain groups of developers~\cite{nam2024,nguyen2024,choudhuri2024HowFarAre}, a thorough understanding of such issues and their underlying causes is needed before we can detect or address them. In \boxStudy{A}, which is currently being conducted, we aim to characterize the relationships between diversity in developers' personal factors, needs, and interaction style. To this end, we have conducted a user study with $n=27$ participants with diverse experience and cognitive profiles. We disregard contextual factors by fixing the experimental context to a code change setting. In a series of progressively harder code change tasks in a single codebase, supported by the programming assistant \textit{Github Copilot Chat}, we combined a thinking aloud setup with post-study semi-structured interviews and questionnaires to get insights into participants' cognitive processes, needs, and challenges they faced. A quantitative analysis of chat messages to Copilot was conducted to identify interaction styles. Currently, these are being related to participants' cognitive and experience profiles to answer \refRQ{I.1}. Preliminary results include distinct needs, interaction styles, and preferences regarding Copilot's support, regarding e.g. the presence of an explanation when generating code, the structure and depth of guidance, and the use of technical terms. These individual differences reinforce the need for tailored support.

There are several existing studies on the effect of contextual factors such as social factors (e.g. support from colleagues, organizations, online communities)~\cite{russo2024NavigatingComplexityGenerative,cheng2024ItWouldWork}, organizational policies~\cite{khojah2024} and cultural values~\cite{lambiase2026InvestigatingRoleCultural} on developers' trust and adoption of programming assistants. However, these studies are relatively high-level, not focusing on how contextual factors shape interaction styles beyond whether or not developers use programming assistants. Moreover, participants are recruited across companies meaning that dynamics within companies and teams may go unnoticed, such as the impact of one developer's tool use on others and whether views on the use of these tools differ between individuals in the same context. In a case study at a software engineering company for \boxStudy{B}, we will address this gap by conducting semi-structured interviews. By mapping how social and organizational context affects needs, attitudes, and interaction styles with programming assistants vary and interconnect between roles and individuals in an organizational context, we will not only answer \refRQ{I.2} but also gain insights for practitioners and managers on how to increase productivity and decrease friction within software engineering teams.

\subsection{Phase II -- Designing Personalization Approaches to Address Diversity}

\boxRQ{II.1}{How can developers' needs be inferred from their interaction style with a programming assistant?}
\boxRQ{II.2}{How do varying developers' needs and interaction styles affect the required support from a programming assistant?}
\boxRQ{II.3}{How can human-in-the-loop methods improve transparency and control of personalization while minimizing developer effort?}

As mentioned earlier in this section, we will explore both implicit and explicit personalization approaches. Implicit personalization reduces effort for users, but creates a risk of profiling by making assumptions about the user~\cite{kirk2024BenefitsRisksBounds}. Explicit personalization decreases this risk through increased transparency and control, but also increases effort. This highlights the needs for carefully designed personalization mechanisms that balances these trade-offs.

Implicit personalization requires our programming assistant to be able to infer developers' needs from their interaction style. In \boxStudy{C}, we will assess if and how we can do this by exploring classical NLP techniques and LLM approaches on our interaction dataset from study \refStudy{A}, answering \refRQ{II.1}. Since we risk increasing effort for specific groups of developers if this mechanism does not function properly~\cite{kirk2024BenefitsRisksBounds}, we will conduct careful testing to ensure our approach is not biased.

Next in \boxStudy{D}, we will employ a vignette survey. In \refStudy{A} and \refStudy{B} we already explore what aspects of interaction should be tailored to individual developers and contexts. For each combination of contextual factors, we will craft several vignettes: scenarios describing the context as well as alternative (non-personalized) ways in which an assistant might support the developer which participants will be asked to rate. By relating contextual factors and preferred forms of support to participants' needs and interaction style, which we obtain through a preliminary questionnaire, we answer \refRQ{II.2}. In this same study, we will also to explore several forms of human-in-the-loop approaches for explicit personalization, such as displaying and allowing edits to inferred needs, and prompting for clarification if a developer's needs are ambiguous, answering \refRQ{II.3}.

\subsection{Phase III -- Implementing a Personalized Programming Assistant Prototype}

\boxRQ{III.1}{How can personalized programming assistance be evaluated?}
\boxRQ{III.2}{How effectively does the proposed prototype personalize interactions?}

In \boxStudy{E}~\cite{richards2025BridgingHCIAI}, we argued how the large scale and high frequency of evaluation that LLM-based conversational assistants require, especially when employing personalization, makes it infeasible to evaluate them only using traditional (manual) human-centered approaches. In this position paper, we advocate combining insights from human-computer interaction (HCI) and artificial intelligence (AI) research to enable human-centered automatic evaluation of LLM-based conversational SE assistants to complement traditional evaluation. By employing LLMs as a simulated user of a programming assistant and through the LLM-as-a-Judge approach, researchers can potentially gain valuable insights aligning with those from actual user studies. Although not empirically answering \refRQ{III.1}, this paper lays the conceptual groundwork for future investigation by identifying requirements for such evaluation and challenges down the road.

In \boxStudy{F}, we will integrate all knowledge synthesized in the previous studies into a personalized programming assistant prototype to evaluate the effectiveness of our personalization approach. This approach encompasses implicit personalization through inference of the developers' needs, and explicit personalization by allowing developers to configure, inspect and edit their needs. We will leverage existing prompting techniques to adapt the assistants' support to developers' needs, guiding the LLM behind the assistant using our previous insights on how this support should be tailored. During the development phase of the prototype, we will implement and employ the automatic evaluation framework described in \refStudy{E}, as well as conduct a pilot study with $n=10$ participants with diverse personal profiles. The pilot study, as well as the later user study, will encompass a range of code-centric tasks including code maintenance, debugging, evolution, testing, and review. The results from the pilot study will be used to inform improvements to the assistant and the automatic evaluation framework. Specific focus will be placed on assessing the presence and extent of bias in the assistant and evaluation framework, which poses significant risks when employing personalization~\cite{kirk2024BenefitsRisksBounds} and automated human-centered evaluation~\cite{richards2025BridgingHCIAI}. Finally, a user study with will be conducted, as well as a large-scale automatic evaluation of the final prototype using our evaluation framework. The results of the user study will be compared with the automatic evaluation to assess the framework's reliability, answering \refRQ{III.1}. Should the framework be shown to be sufficiently informative, reliable, and bias-free, its results will be used together with those of the user study to evaluate the personalized assistant, answering \refRQ{III.2}.

\section{Risks and Mitigation Strategies}
Conducting multiple studies on different aspects of the main research question creates a risk of theoretical drift. We attempt to mitigate this by basing the research in strong theoretical principles (see Section 2).

Conducting several user studies in addition to a case study increases the risk of overgeneralization. This is compounded by personal and contextual factors potentially interacting in complex intersectional ways. Although we limit the factors we address, we cannot fully disentangle these within the scope of this research. We mitigate the risk of overgeneralization by reporting on the diversity in our participant samples and clearly avoiding generalized assumptions about individual differences.

Personalization mechanisms themselves also run the risk of reinforcing bias and stereotyping when designed from overgeneralized insights. We address these concerns by employing human-in-the-loop feedback mechanisms and emphasizing user consent.

\section{Relevance to CHASE Doctoral Symposium}
As research into and adoption of LLMs in software engineering increases rapidly, the need for LLM-based tools to take into account human aspects increases. Since the the research line detailed in this paper is focused on tailoring conversational assistants in order to better support diversity in developers, we think the CHASE Doctoral Symposium provides an ideal opportunity to gain feedback and advice from the community. Besides welcoming any comments in general, we specifically invite the committee to share their views on \begin{enumerate*}[label=\arabic*)]
    \item whether the research is sufficiently based in current literature and relevant theory, and
    \item whether the expected contributions are ambitious enough and make for a strong doctoral thesis.
\end{enumerate*}

\begin{acks}
The author is advised by Prof. Dr. Frits Vaandrager and co-advised by Dr. Mairieli Wessel from Radboud University. 
\end{acks}

\bibliographystyle{ACM-Reference-Format}
\bibliography{references}

\end{document}